\documentstyle[twocolumn,aps]{revtex}

\newtheorem{theorem}{Theorem}
\newtheorem{acknowledgement}[theorem]{Acknowledgement}

\begin{document}
\draft
\title{Magnetic coupling between mesoscopic superconducting rings}
\author{B. J. Baelus*, S. V. Yampolskii%
${{}^\circ}$%
and F. M. Peeters}
\address{Departement Natuurkunde, Universiteit Antwerpen (UIA), Universiteitsplein 1,%
\\
B-2610 Antwerpen, Belgium}
\date{\today}
\maketitle

\begin{abstract}
Using the nonlinear Ginzburg-Landau theory we investigated the dependence of
the magnetic coupling between two concentric mesoscopic superconducting
rings on their thickness. The size of this magnetic coupling increases with
the thickness of the rings.
\end{abstract}

\section{Introduction}

Single mesoscopic superconducting samples have been studied extensively
during the last years. The main part of the studies covered single
superconducting disks (see for example Refs. [1-3]) and rings (see for
example Refs. [4-7]). The properties of single mesoscopic rings of finite
thickness are studied in detail in our previous papers [5]. A single
mesoscopic superconducting ring with a finite thickness tries to expell the
applied magnetic field by inducing supercurrents which create a local
magnetic field opposite to the external one. In some regions the field will
be expelled to the outside, in other regions it will be compressed into the
hole of the ring. As a consequence the total magnetic field is strongly
nonuniform in the neighbourhood of a single superconductor.

What happens if another superconducting ring is placed in the center of such
a ring? The inner ring will try to expell the total nonuniform field instead
of the externally applied one. And this expulsion will influence the outer
ring. In this way, both rings are coupled through the magnetic field. This
coupling will influence the properties of both rings.

Recently, Morelle et al. [8] studied experimentally the interaction between
two concentric superconducting aluminium rings, close to the
superconducting/normal transition. Due to the coupling between the two
rings, they found changes in the $T_{c}(H)$ oscillations of the outer ring.

In the present paper we will investigate theoretically the influence of the
sample thickness on the coupling between two superconducting mesoscopic
rings. With increasing sample thickness the expulsion of the magnetic field
will be more complete, which means that the total field becomes more
nonuniform and the coupling between both rings increases.

\section{Theoretical formalism}

We consider two concentric mesoscopic superconducting rings made of the same
material and with the same thickness d, immersed in a insulating medium and
placed in a perpendicular magnetic field $H_{0}$. The smaller ring has inner
(outer) radius $R_{i}^{\ast }$ ($R_{o}^{\ast }$ ) and the larger ring inner
(outer) radius $R_{i}$ ($R_{o}$). To solve this problem, we expand the
approach for thin superconducting disks of Ref. [2] to a system of two axial
symmetric superconductors. We solve the two coupled Ginzburg-Landau
equations self-consistently 
\begin{equation}
\left( -i\overrightarrow{\nabla }-\overrightarrow{A}\right) ^{2}\Psi =\Psi
\left( 1-\left| \Psi \right| ^{2}\right) \text{ ,}
\end{equation}
\begin{equation}
-\kappa ^{2}\Delta \overrightarrow{A}=\frac{1}{2i}\left( \Psi ^{\ast }%
\overrightarrow{\nabla }\Psi -\Psi \overrightarrow{\nabla }\Psi ^{\ast
}\right) -\left| \Psi \right| ^{2}\overrightarrow{A}\text{ ,}
\end{equation}
where we express all distances in units of the coherence length $\xi $, the
order parameter in $\Psi _{0}=\sqrt{-\alpha /\beta }$ and the vector
potential in $c\hbar /2e\xi $.

The boundary condition for the order parameter is 
\begin{equation}
\left. \overrightarrow{n}\cdot \left( -i\overrightarrow{\nabla }-%
\overrightarrow{A}\right) \Psi \right| _{\rho =R_{i}^{\ast },R_{o}^{\ast
},R_{i},R_{o}}=0\text{ .}
\end{equation}
The boundary condition for the vector potential has to be taken far away
from the sample, where the field equals the applied magnetic field.

The difference between the Gibbs free energy of the superconducting state
and the normal state is determined by the expression 
\begin{equation}
F=\frac{1}{V}\int \left[ 2\left( \overrightarrow{A}-\overrightarrow{A}%
_{0}\right) \cdot \overrightarrow{j}-\left| \Psi \right| ^{4}\right] d\vec{r}%
\text{ ,}
\end{equation}
where the integration is over the total volume $V$ of the superconducting
samples and $\overrightarrow{A}_{0}$ is the vector potential corresponding
to the applied uniform field.

Since we consider sufficiently narrow rings only giant vortex states will
nucleate, which can be characterized by their angular momentum or vorticity.
Therefore, the superconducting states in a double ring system can be denoted
as $(L_{out},L_{in})$, with $L_{out}$ ($L_{in}$) the vorticity of the outer
(inner) ring.

\section{Results}

As an example, we consider a superconducting ring with radii $R_{o}=2.0\xi $
and $R_{i}=1.5\xi $ with a smaller ring in the center with radii $%
R_{o}=2.0\xi $ and $R_{i}^{\ast }=0.6\xi $. Both rings have thus the same
width and we assume that they have the same thickness d and are made of the
same material with $\kappa =0.28$.

Fig. 1 shows the free energy (solid curves) as a function of the applied
magnetic field $H_{0}$ for a double ring configuration with thickness $%
d=0.5\xi $. The free energies of the single rings with the same size as the
inner and the outer ring are given by the thick dashed and the thick dotted
curves, respectively. Notice that everywhere in this paper, the free energy
is expressed in units of $F_{0}=H_{c2}V/8\pi $, where $V$ is the sum of the
volumes of the two rings. This is the reason why the free energy of a single
ring is not equal to $-F_{0}$ at zero magnetic field as it was in, for
example, Ref. [5]. In the single inner ring, superconducting states can
nucleate with vorticities $L_{in}=0$, $1$ and $2$, and in the single outer
ring with $L_{out}=0$ up to $L_{out}=10$. The superconducting/normal
transition fields are $H_{c3}\approx 6.5H_{c2}$ for the inner ring and $%
H_{c3}\approx 6.75H_{c2}$ for the outer ring. The indices $(L_{out},L_{in})$
in Fig. 1 indicate the ground states of the double ring configuration.
Notice that, as compared to the single outer ring, an extra ground state
transition occurs at $H_{0}/H_{c2}\approx 1.5H_{c2}$, where the vorticitity
of the outer ring stays the same, $L_{out}=2$, and the vorticity of the
inner ring changes with one unit, $L_{in}=0\rightarrow 1$. Hence, by putting
an extra ring in the center of a larger ring, the ground state shows extra
transitions. This result corresponds to the experimental result of Morelle
et al. [8]. Notice further that at $H_{0}/H_{c2}\approx 4.3H_{c2}$ both
vorticities change with one unit. The ground state changes from the (6,1)
state to the (7,2) state.

Next we investigate the strength of the interaction between the two rings.
In Fig. 2 we plot the ground state free energy of the coupled rings (solid
curves) and the sum of the ground state free energies of the two
non-interacting single rings (dashed curves) as a function of the applied
magnetic field for three values of the sample thickness: $d/\xi =0.15$, $0.5$%
, and $1.0$. The insets show some crossings in more detail. The difference
between both curves is the interaction energy between the two rings. The
middle curves correspond to the configuration of Fig. 1, i.e. $d/\xi =0.5$.
The interaction is most pronounced for the (2,0), the (5,1) and the (6,1)
state, i.e. at fields just below the transition fields of the inner ring.
Due to this interaction, not only the value of the free energy differs, but
also the transition fields change. For example, the $(1,0)\rightarrow (2,0)$
transition occurs at $H_{0}/H_{c2}\approx 1.01H_{c2}$ neglecting the
coupling and at $H_{0}/Hc2\approx 1.07H_{c2}$ including the coupling.
Moreover, the right inset shows clearly that, neglecting the interaction,
the ground state changes from the (6,1) state into the (6,2) state and then
into the (7,1) state, while including the interaction it transits directly
from the (6,1) state into the (7,2) state. Thus, the coupling between the
two rings leads to the interesting result that the (6,2) state is no longer
a ground state.

It is known that the expulsion of the field from the superconducting rings
increases with increasing sample thickness. This is also shown in Fig. 3,
where we plot the radial distribution of the magnetic field for the (5,1)
state of the double ring configuration of Fig. 1 at $H_{0}/H_{c2}=3.52$ for
three values of the thickness, i.e. $d=0.15\xi $, $0.5\xi $, and $1.0\xi $.
Since the interaction energy is due to the magnetic coupling between both
rings, its value depends strongly on the thickness of the system. This can
be seen from Fig. 2. As compared to the above considered case $(d/\xi =0.5)$
the interaction between the two rings is much smaller for $d/\xi =0.15$,
i.e. the ground state free energy including the coupling is much closer to
the one neglecting the coupling. On the other hand, for $d/\xi =1.0$ the
interaction energy is larger than for $d/\xi =0.5$. Also the number of
ground state transitions of the coupled ring configuration depends on the
sample thickness. For $d/\xi =0.1$5 the (6,2) state is a ground state in the
region $4.27\leq H_{0}/H_{c2}\leq 4.28$, while for $d/\xi =0.5$ and $d/\xi
=1.0$ the ground state changes from the (6,1) state directly to the (7,2)
state.

\section{Conclusions}

We investigated the dependence of the magnetic coupling between two
concentric mesoscopic superconducting rings on the sample thickness. This
coupling results in: 1) the free energy of the double ring configuration is
not exactly the same as the sum of the free energies of the two single
rings. The difference between the two energies, the interaction energy,
increases with increasing sample thickness. 2) For sufficiently thick
samples, some superconducting states are no longer realized as ground states.

\begin{acknowledgement}
\end{acknowledgement}

This work was supported by the Flemish Science Foundation (FWO-Vl), the
''Onderzoeksraad van de Universiteit Antwerpen'', the ''Interuniversity
Poles of Attraction Program - Belgian State, Prime Minister's Office -
Federal Office for Scientific, Technical and Cultural Affairs'', and the
European ESF-Vortex Matter. Discussions with V. V. Moshchalkov are
gratefully acknowledged.

\begin{figure}[tbp]
\caption{The free energy as a function of the applied magnetic field for a
superconducting ring with radii $R_{i}^{\ast }=0.6\protect\xi $ and $%
R_{o}^{\ast }=1.1\protect\xi $ (thick dotted curves) and for a ring with
radii $R_{i}=1.5\protect\xi $ and $R_{o}=2.0\protect\xi $ (thick dashed
curves) and the double ring configuration (thin solid curves). }
\label{Fig1}
\end{figure}

\begin{figure}[tbp]
\caption{The ground state free energy of the double ring configuration of
Fig. 1 (solid curves) and the sum of the ground state free energies of both
single rings (dashed curves) for three values of the sample thickness: $d/%
\protect\xi =0.15$ (upper curves, shifted over $0.2F_{0}$), $d/\protect\xi
=0.5$ (middle curves, shifted over $0.1F_{0}$) and $d/\protect\xi =1.0$
(lower curves) The insets show some crossings in more detail.}
\label{Fig2}
\end{figure}

\begin{figure}[tbp]
\caption{The radial distribution of the magnetic field for the (5,1) state
of the double ring of Fig. 1 at $H_{0}/H_{c2}=3.52$ for three values of the
sample thickness: $d/\protect\xi =0.15$ (solid curves), $d/\protect\xi =0.5$
(dashed curves) and $d/\protect\xi =1.0$ (dotted curves).}
\label{Fig3}
\end{figure}

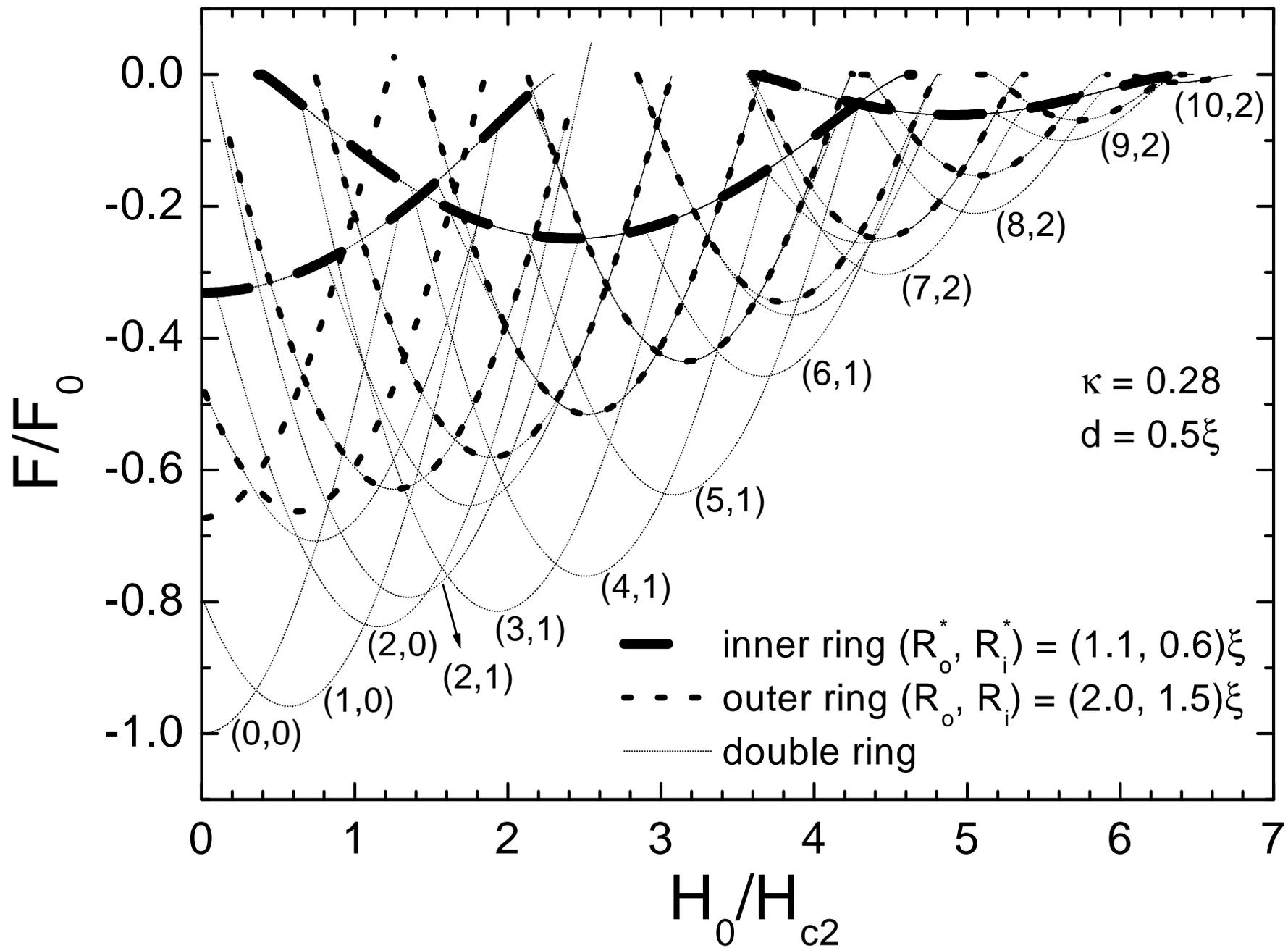
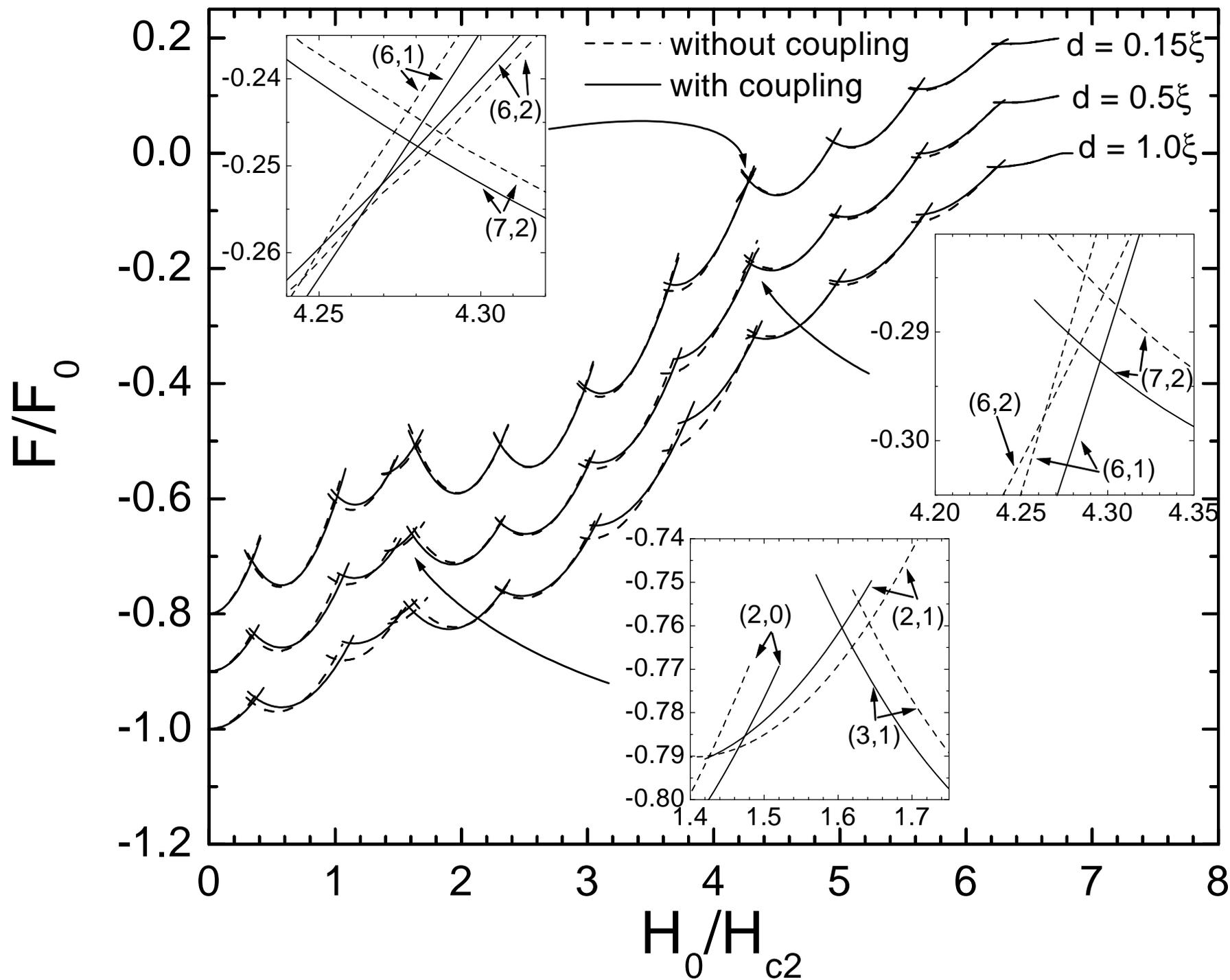
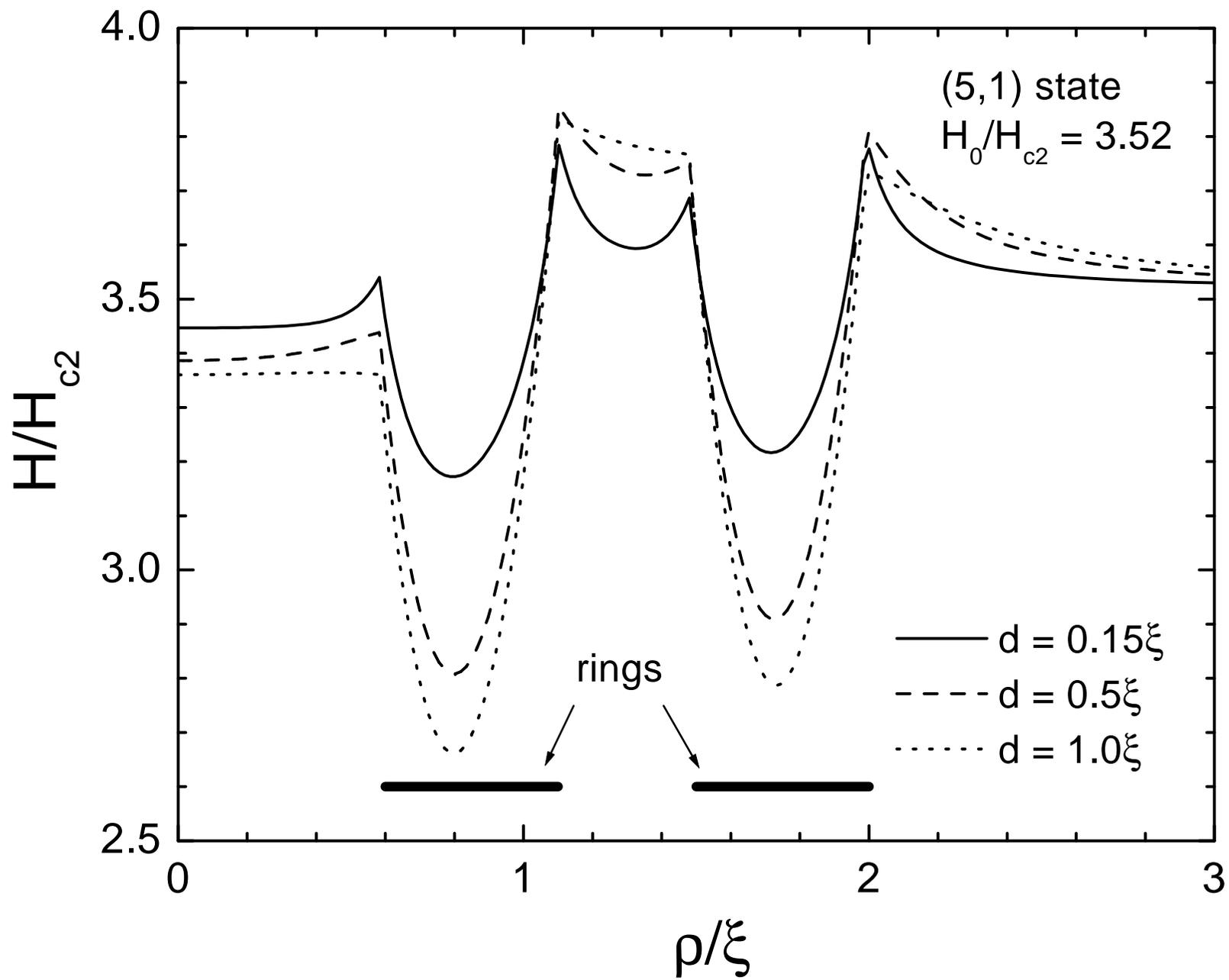
\end{document}